\documentstyle[12pt]{article}

\newcommand{\be}[1]{\begin{equation}\label{#1}}                     
\newcommand{\ba}[1]{\begin{eqnarray}\label{#1}}                     
\newcommand{\ee}{\end{equation}}                                    
\newcommand{\ea}{\end{eqnarray}}                                    
\newcommand{\non}{\nonumber\\\rule{0pt}{30pt}}
\newcommand{\nona}[1]{\nonumber\\\rule{0pt}{#1pt}}
\newcommand{\num}{\\\rule{0pt}{20pt}}

\newcommand{\dis}{\displaystyle}                                    
\newcommand{\Eq}[1]{(\ref{#1})}                                    
\newcommand{\tV}{\widetilde V}                                    
 
\newcommand{\erf}{\mathop{\rm erf}} 
\newcommand{\sign}{\mathop{\rm sign}} 
\newcommand{\stint}{\int_{-\infty}^\infty}                   

\newcommand{\Tr}{\mathop{\rm tr}}
\newcommand{\mm}{\mu}

\newcommand{\freop}{\Bigl(\tilde I+\widetilde V\Bigr)}
\newcommand{\hb}{\hat b}
\newcommand{\ketd}{|0)}
\newcommand{\brad}{(0|}
\makeatletter
\@addtoreset{equation}{section}
\makeatother

\newtheorem{thm}{Theorem}[section]
\newtheorem{lem}[thm]{Lemma}
\begin{document}

\begin{flushright}
December 1997 \\
\end{flushright}

\begin{center}
{\large\bf Asymptotics of the Fredholm
determinant associated with the correlation functions of the quantum
Nonlinear Schr\"odinger equation}

\vspace{2cm}

{\normalsize Nikita Slavnov}\\
\vskip .5em
\vspace{1cm}
{\it Steklov Mathematical Institute,\\
Gubkina 8, 117966,  Moscow, Russia\\}
nslavnov@mi.ras.ru

\end{center}

\vskip4pt

\begin{abstract}
\noindent
The correlation functions of the quantum nonlinear Schr\"odinger equation
can be presented in terms of a Fredholm determinant. The explicit expression
for this determinant is found for the large time and long distance.
\end{abstract}

\section{Introduction}
 In the present paper we derive an asymptotic formula for a Fredholm 
determinant of a linear integral operator of a special type. This problem
arises in connection with calculation of correlation functions 
of integrable models. The example, we consider below, is related to a
correlation function of local fields of the quantum Nonlinear Schr\"odinger
equation out off free fermionic point. Let us remind in brief the
basic definitions of this model.

The  quantum Nonlinear Schr\"odinger equation  can be
described in terms of the canonical Bose fields
$\Psi(x,t),~\Psi^{\dagger}(x,t),~(x,t \in {\bf R})$
obeying the standard commutation relations
\be{Icom}
[\Psi(x,t), \Psi^{\dagger}(y,t)]=\delta(x-y).
\ee
The Hamiltonian of the model is
\be{Hamilton}
{H}={\dis\int \,dx
\left({\partial_x}\Psi^{\dagger}(x)
{\partial_x} \Psi(x)+
c\Psi^{\dagger}(x)\Psi^{\dagger}(x)\Psi(x)\Psi(x)
-h \Psi^{\dagger}(x)\Psi(x)\right).}
\ee
Here $0<c<\infty$ is the coupling constant and $h$
is the chemical potential. The Hamiltonian $H$ acts in the Fock space
with the vacuum vector $|0\rangle$, which
is characterized by the relation:
\be{Ivac}
\Psi(x,t)|0\rangle =0.
\ee
The evolution of the field $\Psi$ with respect to time $t$
is standard
\be{Ievol}
\Psi(x,t)= e^{iHt}\Psi(x,0)e^{-iHt}.
\ee

The quantum Nonlinear Schr\"odinger equation 
describes a one-di\-men\-si\-o\-nal
Bose gas with delta-function interactions. The 
basic thermodynamic  equation of the model
is the Yang--Yang equation \cite{YY} for the energy of an one-particle
excitation $\varepsilon(\lambda)$ in thermal equilibrium
\be{YY}
\varepsilon(\lambda)=\lambda^2-h-\frac{T}{2\pi}
\int_{-\infty}^{\infty}
\frac{2c}{c^2+(\lambda-\mu)^2}\ln
\left(1+e^{-\frac{\varepsilon(\mu)}{T}}
\right)d\mu.
\ee
It is worth mentioning also the integral equation for the total
spectral density of vacancies in the gas $\rho_t(\lambda)$:
\be{Irhot}
2\pi\rho_t(\lambda)=1+\int_{-\infty}^{\infty}
\frac{2c}{c^2+(\lambda-\mu)^2}\vartheta(\mu)
\rho_t(\mu)\,d\mu,
\ee
where
\be{Fermi}
\vartheta(\lambda)=\left(1+\exp\left[
\frac{\varepsilon(\lambda)}{T}\right]\right)^{-1}
\ee
is the Fermi weight. The value $\vartheta(\lambda)\rho_t(\lambda)$
defines the spectral density of particles in the gas.

The time-dependent temperature correlation function
of the local fields is defined by
\be{tempcorrel}
\langle\Psi(0,0)\Psi^\dagger(x,t)\rangle_T=
\frac{\Tr\left( e^{-\frac HT}\psi(0,0)\psi^\dagger(x,t)\right)
}
{\Tr e^{-\frac HT}}.
\ee
Here the trace is taken with respect to all states.

The Fredholm determinant representation for the correlation function
\Eq{tempcorrel} was found in \cite{KKS1}. The classical differential
equations and the Riemann--Hilbert problem, describing the Fredholm
determinant, were formulated in \cite{KKS2}, \cite{KS}. It was shown that the
value of the determinant can be  expressed in terms of solutions
of the Riemann--Hilbert problem or the differential equations, which
in turn can be solved asymptotically for large time and long
distance separation. It is
important that one can find not only the leading term of the asymptotics
of the Fredholm determinant, but also the complete asymptotic expansion.
However, there exists more simple method to calculate at least
the leading term of the asymptotics of the Fredholm determinant. We
would like to emphasize, that because of several reasons, which will
be discussed in the next section, the leading term of the determinant
does not describe the leading term of the asymptotics of the correlation 
function. Therefore, one should know the complete asymptotic expansion.
As we have mentioned already, one can do this via the Riemann--Hilbert
problem and the differential equations methods. Nevertheless, the direct
evaluation of the asymptotics from the Fredholm determinant seems to be 
useful also. First, the direct evaluation can be considered as an
independent confirmation of the results obtained
in the framework of the Riemann--Hilbert problem.
Second, in this case the  calculations become rather simple
and evident so it is clear why the Fredholm determinant can be simplified
for the large distance and time.

\section{The Fredholm determinant}

The determinant representation for the correlation function
\Eq{tempcorrel} is given by the formula
\begin{eqnarray} \langle
\psi(0,0)\psi^\dagger(x,t)\rangle_T= -\frac{e^{-iht}}{2 \pi}
\brad
\frac{\det\left(\tilde{I}+\widetilde{V}\right)}
{\det\left(\tilde{I}-\frac{1}{2\pi}{\widetilde K}_T\right)}
\displaystyle \int_{-\infty}^{\infty}
\hb_{12}(u,v)du dv \ketd.\label{detrep}
\end{eqnarray}
One can find the detail description of all objects entering the
r.h.s. of \Eq{detrep} in \cite{KKS1}--\cite{KS}. Here we
restrict ourselves with only necessary explanation.

We are interested in the large time and long distance asymptotics of
the correlation function.
The Fredholm determinant in the denominator of \Eq{detrep} does
not depend on time $t$ and distance $x$, therefore it can be considered
as some constant. The important objects, which possess nontrivial
dependency on $x$ and $t$, are the factor $\hb_{12}(u,v)$ and the
Fredholm determinant $\det\freop$. Both of them depend also on auxiliary
quantum operators---dual fields. Let us describe, first of all, these
auxiliary operators.

The dual fields  $\psi(\lambda)$,
$\phi_{D}(\lambda)$ and $\phi_{A}(\lambda)$ (originally the last two
fields were denoted as $\phi_{D_1}(\lambda)$ and $\phi_{A_2}(\lambda)$)
were introduced in \cite{KKS1} in order to remove two-body
scattering and to reduce the model to free fermionic one.
These fields act in an auxiliary Fock space  having vacuum vector $\ketd$ and
dual vector $\brad$ (one should not confuse the auxiliary vacuum vector
$\ketd$ with the vector $|0\rangle$ in \Eq{Ivac}).
Each of these fields can be presented as a sum of creation and annihilation parts
\be{dualfields}
\begin{array}{rcl}
\phi_{A}(\lambda)&=&q_{A}(\lambda)+p_{D}(\lambda),\\
\phi_{D}(\lambda)&=&q_{D}(\lambda)+p_{A}(\lambda),\\
\psi(\lambda)&=&q_\psi(\lambda)+p_\psi(\lambda).
\end{array}
\ee
Here $p(\lambda)$ are annihilation parts of dual fields:
$p(\lambda)\ketd=0$; $q(\lambda)$ are creation parts of dual fields:
$\brad q(\lambda)=0$.  

The only nonzero commutation relations are
\be{commutators}
\begin{array}{l}
{}[p_{A}(\lambda),q_\psi(\mu)]=
[p_\psi(\lambda),q_{A}(\mu)]=\ln h(\mu,\lambda),\num
{}[p_{D}(\lambda),q_\psi(\mu)]=
[p_\psi(\lambda),q_{D}(\mu)]=\ln h(\lambda,\mu),\num
{}[p_\psi(\lambda),q_\psi(\mu)]=\ln [h(\lambda,\mu)h(\mu,\lambda)],
\quad\mbox{where}\quad
 {\dis h(\lambda,\mu)=\frac{\lambda-\mu+ic}{ic}.}
\end{array}
\ee
Recall that $c$ is the coupling constant in \Eq{Hamilton}.
It follows immediately from \Eq{commutators} that the dual fields
belong to an Abelian sub-algebra
\be{Abel} [\psi(\lambda),\psi(\mu)]=
[\psi(\lambda),\phi_a(\mu)]=
[\phi_b(\lambda),\phi_a(\mu)]=0,
\ee
where $a,b=A,D$. The properties \Eq{commutators}, \Eq{Abel}, in fact,
permit us to treat the dual fields as complex functions,
which are holomorphic in some neighborhood of the real axis.

From this point of view one can consider the factor $\hb_{12}(u,v)$ and
the determinant $\det\freop$ in \Eq{detrep} as they depend on some
functional parameters $\psi(\lambda)$, $\phi_A(\lambda)$ and
$\phi_D(\lambda)$. With rather smooth restrictions (like analyticity
of these functional parameters in a neighborhood of the real axis) one
can find an asymptotic expansion of the factor,
as well as the Fredholm determinant,
for the large $x$ and $t$ separation. The terms of this asymptotic series
functionally depend on dual fields. However, in order to find the asymptotics
for the correlation function in the l.h.s. of \Eq{detrep} one has to calculate
the vacuum expectation value of the expression obtained. The detailed 
analysis of the Riemann--Hilbert problem and the differential equations
shows that some of the terms of the asymptotic series, containing negative 
powers of $t$ (or $x$), provide a non-trivial contribution into
the leading term of the asymptotics after the calculation of their
 vacuum expectation value.
That is why, in order to describe the asymptotic behavior of the correlation
function, one must find not only the leading term of the series, depending on
dual fields, but some corrections to this term also.

In the present paper we are not going to calculate the 
asymptotics of the correlation function. Our main goal is to
illustrate the principal possibility of a simplification of the
Fredholm determinant $\det\freop$ in the regime of large time and
long distance. Therefore, in particular, we restrict ourselves with the
calculation of the leading term and the first important correction of the
determinant only. The factor $\hb_{12}$ remains out off framework
of our consideration. In order to describe this factor (as well as to find
more precise estimates for the determinant) one has to apply more powerful
(but not so evident!) methods of the Riemann--Hilbert problem
and the differential equations.

The integral operator $\tilde I+\tV$, which Fredholm determinant enters
 the r.h.s. of (\ref{detrep}), acts on the real axis as
\begin{eqnarray}
\left(\tilde{I}+\widetilde{V}\right)\circ f(\mu)
=f(\lambda)+\displaystyle \int_{-\infty}^{\infty}\tV(\lambda,\mu)
f(\mu)d\mu,
\end{eqnarray}
 where $f(\lambda)$ is some trial function.
Thus, $\tilde I$ is the identity operator.

The kernel $\tV(\lambda,\mu)$ can be presented in the form
\be{FDform}
\tV(\lambda,\mu)=\frac 1{\lambda-\mu}
\stint\,du(E_+(\lambda|u)E_-(\mu|u)-E_-(\lambda|u)E_+(\mu|u)),
\ee
Here functions $E_\pm$ introduced in \cite{KKS2} are equal to:
\ba{eplus}
&&{\dis
E_+(\lambda|u)=\frac{1}{2\pi}\frac{Z(u,\lambda)}{Z(u,u)}
\left(\frac{e^{-\phi_{A}(u)}}{u-\lambda+i0}
+\frac{e^{-\phi_{D}(u)}}{u-\lambda-i0}\right)
\sqrt{\vartheta(\lambda)}}\nona{19}
&&\hskip3cm{\dis \times
e^{\psi(u)+\tau(u)+\frac12(
\phi_{D}(\lambda)+\phi_{A}(\lambda)-\psi(\lambda)-\tau(\lambda))},}
\ea
\hskip1mm
\be{eminus}
E_-(\lambda|u)=\frac{1}{2\pi}Z(u,\lambda)
e^{\frac12(\phi_{D}(\lambda)+\phi_{A}(\lambda)
-\psi(\lambda)-\tau(\lambda))}\sqrt{\vartheta(\lambda)},
\ee
where the function $Z(\lambda,\mu)$ is defined by
\be{Z}
Z(\lambda,\mm)=\frac{e^{-\phi_{D}(\lambda)}}{h(\mm,\lambda)}+
\frac{e^{-\phi_{A}(\lambda)}}{h(\lambda,\mm)}.
\ee
Here $\psi(\lambda)$, $\phi_{D}(\lambda)$ and $\phi_{A}(\lambda)$
are just the dual fields \Eq{dualfields}.

Recall also that the function $\vartheta(\lambda)$ is the Fermi weight
\Eq{Fermi}. This function defines the dependence of the correlation function on
temperature and chemical potential. The function $\tau(\lambda)$ is the only
function depending on time and distance:
\be{tau}
\tau(\lambda)=it\lambda^2-ix\lambda.
\ee
Later on we shall study the asymptotics of the determinant \Eq{tempcorrel}
in the limit $t\to\infty$, $x\to\infty$, however the ratio $x/2t=\lambda_0$ 
remains fixed. Therefore, it is convenient to rewrite the function $\tau(\lambda)$
in terms of variables $t$ and $\lambda_0$:
\be{tau1}
\tau(\lambda)=
it(\lambda-\lambda_0)^2-it\lambda_0^2.
\ee
Thus, we have described the Fredholm determinant in the
r.h.s. of (\ref{detrep}). The  auxiliary quantum
operators---dual fields---enter the kernel $\tV$.
However, due to the property \Eq{Abel} the Fredholm determinant is
well defined. It is important that  the kernel $\tV(\lambda,\mu)$ 
possesses no singularity at $\lambda=\mu$.

\section{Properties of the integral kernels}

For the investigation of the asymptotics of the Fredholm determinant
we use well known representation
\be{rep}
\det(I+\tV)
=\exp\left\{
\sum_{n=1}^\infty\frac{(-1)^n}{n}\int 
\tV(\lambda_1,\lambda_2)\tV(\lambda_2,\lambda_3)
\dots \tV(\lambda_n,\lambda_1)\,d^n\lambda
\right\}
\ee

Recall, that we consider the case, when $t\to\infty$, $x\to\infty$ and
the ratio $x/2t=\lambda_0$ remains fixed. In this case the original
kernel $\tV(\lambda,\mu)$ can be simplified. Indeed,
for the large $t$ one can take the integral in \Eq{FDform} with respect to
$u$ (see \Eq{asymptII}). Up to terms of the order $t^{-1/2+\varepsilon}$, where
$\varepsilon$ is arbitrary positive, we have
\be{mainkernel1}
\tV(\lambda,\mu)=\frac{e^{\frac{1}{2}(\tau(\lambda)-\tau(\mu))}
S_+(\lambda,\mu)+
e^{\frac{1}{2}(\tau(\mu)-\tau(\lambda))}
S_-(\lambda,\mu)}{\lambda-\mu},
\ee
with
\ba{mSplus}
{\dis S_+(\lambda,\mu)}&=&{\dis
\frac{i}{2\pi}e^{\frac{1}{2}\biggl(\psi(\lambda)-\psi(\mu)
+\phi_{D}(\lambda)+\phi_{A}(\lambda)
+\phi_{D}(\mu)+\phi_{A}(\mu)\biggr)}
Z(\lambda,\mu)}\non
&\hspace{-1.5cm}\times&{\dis\hspace{-1cm}
\left(e^{-\phi_{D}(\lambda)}\theta(\lambda-\lambda_0)-
e^{-\phi_{A}(\lambda)}\theta(\lambda_0-\lambda)\right)
\sqrt{\vartheta(\lambda)}\sqrt{\vartheta(\mu)}},
\ea
\ba{mSminus}
{\dis S_-(\lambda,\mu)}&=&{\dis
-\frac{i}{2\pi}e^{\frac{1}{2}\biggl(\psi(\mu)-\psi(\lambda)
+\phi_{D}(\lambda)+\phi_{A}(\lambda)
+\phi_{D}(\mu)+\phi_{A}(\mu)\biggr)}
Z(\mu,\lambda)}\non
&\hspace{-1.5cm}\times&{\dis\hspace{-1cm}
\left(e^{-\phi_{D}(\mu)}\theta(\mu-\lambda_0)-
e^{-\phi_{A}(\mu)}\theta(\lambda_0-\mu)\right)
\sqrt{\vartheta(\lambda)}\sqrt{\vartheta(\mu)}}.
\ea

Let us consider some properties of integral
operators possessing the kernels of  type \Eq{mainkernel1}
\be{kernel}
V(\lambda,\mu)=\frac{e^{\frac{1}{2}(\tau(\lambda)-\tau(\mu))}
S_+(\lambda,\mu)+
e^{\frac{1}{2}(\tau(\mu)-\tau(\lambda))}
S_-(\lambda,\mu)}{\lambda-\mu}.
\ee
Here $S_\pm(\lambda,\mu)$ are some functions 
(in particular \Eq{mSplus} and \Eq{mSminus}), which assumed to
be integrable at the real axis. These 
functions may depend on the ratio $\lambda_0=x/2t$, 
but not on $x$ and $t$ separately. We also demand that the
kernel $V(\lambda,\mu)$ possesses no singularity at the point
$\lambda=\mu$, i. e. the functions $S_\pm$ are smooth in
the vicinity of $\lambda=\mu$ and
\be{nosing}
S_+(\lambda,\lambda)=-S_-(\lambda,\lambda),
\ee

The main  goal of this section is to study the
 operator product of the kernels \Eq{kernel}. 
In what follows we shall drop out all the terms,
vanishing in the limit $t\to\infty$.

The following simple lemma will play an important role.

\begin{lem}~~{\it Let two kernels $V^{(j)}(\lambda,\mu),
\quad j=1,2$ of type \Eq{kernel} are given
\be{kernelj}
V^{(j)}(\lambda,\mu)=\frac{e^{\frac{1}{2}(\tau(\lambda)-\tau(\mu))}
S_+^{(j)}(\lambda,\mu)+
e^{\frac{1}{2}(\tau(\mu)-\tau(\lambda))}
S_-^{(j)}(\lambda,\mu)}{\lambda-\mu},\qquad j=1,2,
\ee
with
\be{nosingj}
S_+^{(j)}(\lambda,\lambda)=-S_-^{(j)}(\lambda,\lambda),\qquad j=1,2.
\ee
Then the kernel   $V^{(12)}(\lambda,\mu)$ 
\be{prodkern}
V^{(12)}(\lambda,\mu)=
\stint V^{(1)}(\lambda,\nu)V^{(2)}(\nu,\mu)\,d\nu,
\ee
is equal to
\be{kernel12}
V^{(12)}(\lambda,\mu)=
{\dis\frac{e^{\frac{1}{2}(\tau(\lambda)-\tau(\mu))}
S_+^{(12)}(\lambda,\mu)+
e^{\frac{1}{2}(\tau(\mu)-\tau(\lambda))}
S_-^{(12)}(\lambda,\mu)}{\lambda-\mu}},
\ee
and
\be{nosing12}
S_+^{(12)}(\lambda,\lambda)=-S_-^{(12)}(\lambda,\lambda).
\ee
}
\end{lem}
{\sl Proof.}~~Let us notice that due to
\Eq{nosingj} one can replace  the denominator
$\lambda-\mu$ in \Eq{kernelj}
by, for example, $\lambda-\mu-i0$. Then, applying formul\ae~\Eq{asymptII}
of Appendix A, we immediately arrive at the representation \Eq{kernel12}
for $V^{(12)}(\lambda,\mu)$ with
\ba{Splus}
&&{\dis S_+^{(12)}(\lambda,\mu)=
\stint \,d\nu S_+^{(1)}(\lambda,\nu)S_+^{(2)}(\nu,\mu)}\non
&&{\dis
\times\left(\frac{\theta(\nu-\lambda_0)}{\nu-\mu-i0}
+\frac{\theta(\lambda_0-\nu)}{\nu-\mu+i0}
-\frac{\theta(\nu-\lambda_0)}{\nu-\lambda+i0}
-\frac{\theta(\lambda_0-\nu)}{\nu-\lambda-i0}\right),}
\ea
\ba{Sminus}
&&{\dis S_-^{(12)}(\lambda,\mu)=
\stint \,d\nu S_-^{(1)}(\lambda,\nu)S_-^{(2)}(\nu,\mu)}\non
&&{\dis
\times\left(\frac{\theta(\nu-\lambda_0)}{\nu-\mu+i0}
+\frac{\theta(\lambda_0-\nu)}{\nu-\mu-i0}
-\frac{\theta(\nu-\lambda_0)}{\nu-\lambda-i0}
-\frac{\theta(\lambda_0-\nu)}{\nu-\lambda+i0}\right).}
\ea
where $\theta(\lambda)$ is the step function.
Obviously,
\be{cond}
S_+^{(12)}(\lambda,\lambda)=
2\pi i\sign(\lambda-\lambda_0)
S_+^{(1)}(\lambda,\lambda)S_+^{(2)}(\lambda,\lambda)=
-S_-^{(12)}(\lambda,\lambda),
\ee
where the sign function as usual is equal to
\be{sign}
\sign(\lambda)=\theta(\lambda)-\theta(-\lambda).
\ee
The lemma is proved.

The direct corollary of the lemma is that any power of the
kernel \Eq{kernel} 
\be{power0}
V^n(\lambda,\mu)=\int V(\lambda,\lambda_1)V(\lambda_1,\lambda_2)
\dots V(\lambda_{n-1},\mu)\,d^{n-1}\lambda
\ee
has just the same structure as the kernel $V(\lambda,\mu)$:
\be{power}
\biggl(V(\lambda,\mu)\biggr)^n=
\frac{e^{\frac{1}{2}(\tau(\lambda)-\tau(\mu))}
S_{+,n}(\lambda,\mu)+
e^{\frac{1}{2}(\tau(\mu)-\tau(\lambda))}
S_{-,n}(\lambda,\mu)}{\lambda-\mu},
\ee
where we have set $S_{\pm,1}(\lambda,\mu)\equiv S_{\pm}(\lambda,\mu)$.
Recall that we drop out all the terms of the order $t^{-1/2}$.

A very simple formula for $S_{\pm,n}(\lambda,\lambda)$ follows from
\Eq{cond}. Indeed, we have
\be{prelim}
S_{\pm,n}(\lambda,\lambda)=2\pi i\sign(\lambda-\lambda_0)
S_{\pm,n-1}(\lambda,\lambda)S_{\pm,1}(\lambda,\lambda),
\ee
and hence,
\be{powerSll}
S_{\pm,n}(\lambda,\lambda)=\biggl(2\pi i\sign(\lambda-\lambda_0)
\biggr)^{n-1}\biggl(S_{\pm}(\lambda,\lambda)\biggr)^n.
\ee
%


The traces of powers of $V$ can be described via formula \Eq{power}
\ba{trpower}
\Tr\biggl(V(\lambda,\mu)\biggr)^n&=&{\dis
\stint\tau'(\lambda)S_{+,n}(\lambda,\lambda)\,d\lambda}\non
&-&{\dis
\stint 
\left.\frac{\partial}{\partial\mu}\biggl(S_{+,n}(\lambda,\mu)+
S_{-,n}(\lambda,\mu)\biggr)\right|_{\mu=\lambda}\,d\lambda.}
\ea
The second term in \Eq{trpower} does not depend on $t$ (it depends only
on  $\lambda_0$). The first term is proportional to
$t$ and, hence, just this term describes asymptotic of determinant.
Due to \Eq{powerSll} we have
\be{astrpower}
\Tr\biggl(V(\lambda,\mu)\biggr)^n\longrightarrow
\stint\tau'(\lambda)\biggl(2\pi 
i\sign(\lambda-\lambda_0)\biggr)^{n-1} 
\biggl(S_{+}(\lambda,\lambda)\biggr)^n\,d\lambda.
\ee
Substituting this estimate into \Eq{rep} we obtain
\ba{prelasdet}
\ln\det(\tilde I+V)&\longrightarrow&{\dis\frac{1}{2\pi i}\stint
\tau'(\lambda)\sign(\lambda-\lambda_0)}\non
&\times&{\dis\ln\Bigl(1+
2\pi i\sign(\lambda-\lambda_0)S_+(\lambda,\lambda)\Bigr)\,d\lambda
.}
\ea
It is sufficient now to substitute the explicit expression \Eq{mSplus} into
the last formula. Clearly that
\be{tauprim}
\tau'(\lambda)=2it(\lambda-\lambda_0),
\ee
\be{Splusll}
S_+(\lambda,\lambda)=\frac{i\vartheta(\lambda)}{2\pi}
\sign(\lambda-\lambda_0)\left(1+
e^{\bigl(\phi_{A}(\lambda)-\phi_{D}(\lambda)\bigr)
\sign(\lambda-\lambda_0)}\right).
\ee
Therefore we obtain for the leading term of the asymptotics of the Fredholm
determinant \Eq{detrep}
\ba{asdet}
&&\ln\det(\tilde I+\tV){\longrightarrow}{\dis\frac{t}{\pi}\stint
|\lambda_0-\lambda|}\non
&&{\dis\hspace{-1cm}\times
\ln\Bigl\{1-\vartheta(\lambda)
\biggl(1+e^{\bigl(\phi_{A_2}(\lambda)-\phi_{D_1}(\lambda)\bigr)
\sign(\lambda-\lambda_0)}\biggr)\Bigr\}
\,d\lambda
+O(1).}
\ea

Thus, we have obtained the leading term for the asymptotics
of the Fredholm determinant.
It is interesting to consider the
free fermionic limit of this result. In this case
the coupling constant $c$ goes to infinity, and all
commutation relations between annihilation and creation parts of dual
fields vanish. Thus, dual fields do not contribute into the vacuum
expectation value and, hence,  one can put them equal to zero. 
The equation \Eq{YY} can be solved explicitly, so we find for the
Fermi weight
\be{FreeFer}
\vartheta(\lambda)=\left(e^{\frac{\lambda^2-h}T}+1\right)^{-1},
\qquad c\to\infty.
\ee
Thus, we arrive at
\be{asdetff}
\ln\det(\tilde I+\tV)\stackrel{c\to\infty}
{\longrightarrow}\frac{1}{2\pi}\stint
|x-2\lambda t|\ln\left(\frac{
e^{\frac{\lambda^2-h}T}-1}
{e^{\frac{\lambda^2-h}T}+1}\right)
\,d\lambda,
\ee
which exactly coincides with the result obtained in the \cite{IIKV}.
\section{Improved formula for the asymptotics}

The leading term of the asymptotics of the Fredholm determinant is given by
the formula \Eq{asdet}. However, this formula is not complete due to the 
presence of dual fields and it does not provide the correct result
for the asymptotics of the correlation function.
Indeed, if we calculate the vacuum expectation of $\det(\tilde I+\tV)$ 
defined by \Eq{asdet}, we obtain the result, which is very similar to
the free fermionic one:
\ba{attempt}
&&\hspace{-0.8cm}(0|{\dis\exp\left\{\frac{1}{2\pi}\stint
|x-2\lambda t|
\ln\Bigl\{1-\vartheta(\lambda)
\biggl(1+e^{\bigl(\phi_{A}(\lambda)-\phi_{D}(\lambda)\bigr)
\sign(\lambda-\lambda_0)}\biggr)\Bigr\}
\,d\lambda
\right\}}|0)\non
&&\hspace{3cm}={\dis
\exp\left\{\frac{1}{2\pi}\stint
|x-2\lambda t|\ln\bigl\{1-
2\vartheta(\lambda)\bigr\}
\,d\lambda
\right\}}.
\ea
Here we have used the fact that
dual fields $\phi_{D}(\lambda)$ and $\phi_{A}(\lambda)$ do
not contain non-commutative operators. 
However, it is easy to show that some corrections to the leading
term strongly change the situation. Let us find one of these corrections.
Namely, it is not difficult to find, how the Fredholm determinant
depends on dual field $\psi(\lambda)$. To do it, let us pay attention
to the fact that $\psi(\lambda)$ enters the kernel $\tV(\lambda,\mu)$ 
\Eq{mainkernel1}  only in the combination with $\tau(\lambda)$: $\quad
\psi(\lambda)+\tau(\lambda)$. More explicitly
\be{Spmtilde}
S_\pm(\lambda,\mu)=e^{\pm\frac{1}{2}(\psi(\lambda)-
\psi(\mu))}\tilde S_\pm(\lambda,\mu),
\ee
where $\tilde S_\pm(\lambda,\mu)$ do not depend on $\psi(\lambda)$.
One can rewrite now formul\ae~\Eq{trpower}, \Eq{astrpower} as
\ba{trpower1}
\Tr\biggl(\tV(\lambda,\mu)\biggr)^n&=&{\dis
\stint(\tau'(\lambda)+\psi'(\lambda))
\tilde 
S_{+,n}(\lambda,\lambda)\,d\lambda}\non
&-&{\dis
\stint 
\left.\frac{\partial}{\partial\mu}\biggl(\tilde S_{+,n}(\lambda,\mu)+
\tilde S_{-,n}(\lambda,\mu)\biggr)\right|_{\mu=\lambda}\,d\lambda,}
\ea
and
\ba{astrpower1}
&&\Tr\biggl(\tV(\lambda,\mu)\biggr)^n\longrightarrow
C_0^{(n)}\non
&&{\dis\hspace{-6mm}+
\stint\biggl(\tau'(\lambda)+\psi'(\lambda)\biggr)\biggl(2\pi 
i\sign(\lambda-\lambda_0)\biggr)^{n-1} 
\biggl(\tilde S_{+}(\lambda,\lambda)\biggr)^n\,d\lambda,}
\ea
where $C_0^{(n)}$ depend only on dual fields $\phi_{D}(\lambda)$ and 
$\phi_{A}(\lambda)$. Thus, we arrive at the improved formula for 
the asymptotics of the Fredholm determinant
\ba{asdet1}
&&\hspace{-6mm}\det(\tilde I+\tV){\longrightarrow}{\dis 
C(\phi_{D},\phi_{A})\exp\left\{
\frac{1}{2\pi}\stint\biggl(|x-2\lambda 
t|-i\sign(\lambda-\lambda_0)\psi'(\lambda)\biggr)\right.}\non 
&&\hspace{2cm}\times{\dis\left.
\ln\Bigl\{1-\vartheta(\lambda)
\biggl(1+e^{\bigl(\phi_{A}(\lambda)-\phi_{D}(\lambda)\bigr)
\sign(\lambda-\lambda_0)}\biggr)\Bigr\}
\,d\lambda
\right\}.}
\ea
Now the r.h.s. of \Eq{asdet1} contains non-commutative operators
(in spite of all dual fields still commute with each other), and 
its vacuum expectation value becomes very nontrivial. The method of the
calculation of the vacuum expectation of the expression \Eq{asdet1} was
developed in \cite{KS1}. It was shown that the result can be expressed
in terms of a solution of a nonlinear integral equation, closely
related to the Thermodynamic Bethe Ansatz equations.

\section*{Acknowledgments}
This work was supported by 
the Russian Foundation of Basic Research under
Grant No. 96-01-00344.

\appendix

\section{Asymptotics of singular integrals}
Consider the asymptotics of the integral 
\be{I}
I(\lambda,\lambda_0,t,[\varphi])=\stint\frac{e^{\tau(u)}}{u-\lambda-i0}
\varphi(u,\lambda)\,du.
\ee
where
\be{Atau}
\tau(u)=it(u-\lambda_0)^2-it\lambda_0^2.
\ee
Recall, that $ t\to+\infty$, while $\lambda_0$ is fixed. The parameter
$\lambda$ is arbitrary real, the function $\varphi(u,\lambda)$ is
holomorphic with respect to $u$ in some neighborhood of the real axis.

Using the steepest descent method one can estimate the integral \Eq{I}
\be{present} 
I(\lambda,\lambda_0,t,[\varphi])=\varphi(\lambda,\lambda)
I_1(\lambda,x,t)+O(t^{-1/2}),
\ee
where
\be{I2}
I_1(\lambda,\lambda_0,t)=
\stint\frac{e^{\tau(u)}}{u-\lambda-i0}\,du.
\ee

The last integral can be expressed in terms of the ``error function''
\be{Avalint}
I_1(\lambda,\lambda_0,t)=i\pi e^{\tau(\lambda)}
\left[\erf\left((\lambda-\lambda_0)\sqrt t e^{-i\frac\pi4}\right)
+1\right],
\ee
where, by definition
\be{Adeferr}
\erf(x)=\frac2{\sqrt\pi}\int_0^xe^{-z^2}\,dz.
\ee
Clearly, that the r.h.s. of \Eq{Avalint} possesses no uniform asymptotic with 
respect to $t$ for all $\lambda$. Its asymptotic behavior strongly
depends of position of $\lambda$ with respect to the $\lambda_0$. Indeed,
if, for example, $\lambda-\lambda_0\gg t^{-1/2}$, then the argument of the
error function goes to plus infinity, and we have

\be{asymptI1}
I_1(\lambda,\lambda_0,t)\to 2\pi ie^{\tau(\lambda)}
\qquad \mbox{for} \qquad
\lambda-\lambda_0\gg t^{-1/2}.
\ee
However, if $\lambda_0-\lambda\gg t^{-1/2}$, then
\be{asymptI2}
I_1(\lambda,\lambda_0,t)\to 0
\qquad  \mbox{for} \qquad
\lambda_0-\lambda\gg t^{-1/2}.
\ee
Finally, if the difference $\lambda-\lambda_0$
is of the order of $t^{-1/2}$, then we can not apply asymptotic formul\ae~
for the error function. Nevertheless, it is easy to see, that for
arbitrary smooth, integrable at the real axis function $f(\lambda)$ the
following estimate holds
\be{Aestweak}
\stint I_1(\lambda,\lambda_0,t) f(\lambda)\,d\lambda=
2\pi i\int_{\lambda_0}^\infty e^{\tau(\lambda)}f(\lambda)\,d\lambda
+ O(t^{-1/2}).
\ee
Thus, one can say, that in a weak sense the original integral
$I(\lambda,\lambda_0,t,[\varphi])$ possesses the
following asymptotics for the large $t$
\be{asymptI11}
I(\lambda,x,t,[\varphi])=
2\pi i e^{\tau(\lambda)}\varphi(\lambda,\lambda)
\theta(\lambda-\lambda_0)+
O(t^{-1/2+\varepsilon}),
\ee
where $\varepsilon$ is arbitrary positive, and
$\theta(\lambda)$ is  a step function:
\be{theta}
\theta(\lambda)=
\left\{\begin{array}{c}
1,\qquad \lambda>0,\\
0,\qquad \lambda<0.
\end{array}
\right.
\ee

Using the same way we can find asymptotic formul\ae~for similar 
singular integrals. Below the list of formul\ae, which were used in 
the section 3, is given.  
\be{asymptII} 
\begin{array}{l}
{\dis\stint\frac{e^{\pm\tau(u)}}{u-\lambda\mp i0}
\varphi(u,\lambda)\,du=\pm
2\pi i e^{\pm\tau(\lambda)}\varphi(\lambda,\lambda)
\theta(\lambda-\lambda_0)+
O(t^{-1/2+\varepsilon})},\non
{\dis\stint\frac{e^{\pm\tau(u)}}{u-\lambda\pm i0}
\varphi(u,\lambda)\,du=
\mp 2\pi i e^{\pm\tau(\lambda)}\varphi(\lambda,\lambda)
\theta(\lambda_0-\lambda)+
O(t^{-1/2+\varepsilon})}.
\end{array}
\ee

\end{document}